# Aging Concept in Population Dynamics

Kazumi Suematsu

Institute of Mathematical Science

Ohkadai 2-31-9, Yokkaichi, Mie 512-1216, JAPAN

Fax: +81 (0) 593 26 8052;  E-mail: suematsu@m3.cty-net.ne.jp

## Summary

Author's early work on aging is developed to yield a relationship between life spans and the velocity of aging. The mathematical analysis shows that the mean extent of the advancement of aging throughout one's life is conserved, or equivalently, the product of the mean life span, $T$, and the mean rate of aging, $\bar{v}_A$, is constant, $T \times \bar{v}_A = k$. The result is in harmony with our experiences: It accounts for the unlimited replicability of tumor cells, and predicts the prolonged life spans of hibernating hamsters according to the equation, $T \cong (k/\bar{v}_{AN}) 1/(1-\chi)$ (the subscript N denotes non-hibernation and $\chi$ the fraction of hibernation), in accordance with the Lyman and coworkers experiment. Comparing the present result and the empirical relationship between life spans of various mammals and basal metabolic rates, it is suggested that the mean rate of aging is intimately connected with the mean basal metabolic rate. With the help of this information, we inquire the reason of the difference in mean life spans between women and men, the result showing that the relative mean life span of women to men is $\cong 1.08$ for various nations, which is close to the corresponding relative value of the basal metabolic rate. The present theory suggests, however, that this relationship between life spans and basal metabolic rates must be treated with caution.

## Key Words

advancement of aging/ velocity of aging/ hibernation/ relative mean life-span/ relative metabolism/

## §1. Introduction

Aging is the oldest subject in human science. In spite of long and strong concern to aging phenomena, to date human cultures have not laid gerontology on a firm foundation of science. The reason may come from the extreme complexity of an organism composed of a vast variety of molecular species, infinitely diversified functions and structures together with turn over mechanisms, which has rejected so far the traditional microscopic approach based on the Newtonian mechanics. This circumstance would be sufficient to discourage most theorists. Thus, it is natural that modern theorists who think much of mathematical rigor didn't take up seriously aging as a research subject. Under these circumstances, a great deal of observed data have been accumulated by experimental biologists which in turn induced countless conjectures about the nature of aging. An essential feature is that most of them remain qualitative and explanatory arguments (Sacher, 1967, 1980, 1982; Medvedev, 1990). As a consequence, those propositions fail to predict new phenomena and new experiments. Thus far aging research has not yet been quantitative science.

What is strange with aging is that it appears that the more deeply one analyzes an aged body, the more obscure the aging process becomes; for instance, we can correctly guess one's age from his appearance, thus aging is recognizable in human level, while it is hard to guess one's age by observing his single cell, and almost impossible by a single molecule, a single atom and so forth. It appears that aging is a phenomenon characteristic of macroscopic systems; in microscopic levels, aging seems not to exist in the usual meaning. For this reason, we have pursued so far a macroscopic approach of aging. Not surprisingly, it is found that we can say many about the nature of aging without entering the complication of the biological machinery.

According to Strehler, there is one effect of aging which is accurately and quantitatively studied in light of the rigor of modern science. That is the probability of survival, $P$, as a function of age. It has been widely known that the survival curves in advanced societies have a common shape, the plateau in youth, the rapidly descending zone of middle age and the exponentially decreasing zone at higher age, so that one can correctly estimate the death probability, $Q = 1 - P$, at an arbitrary age (Strehler, 1967). Thus it will be reasonable for us to attempt to construct an aging theory on the basis of the well-established quantity, $P$.

In this report, we push forward our arguments on the basis of the relationship, $A = -k \log P$, derived earlier (Suematsu, 1995), between the extent of the advancement of aging, $A$, and the probability of survival, $P$. We show that this relation leads us to an interesting consequence: the total extent of the advancement of aging throughout one's life is conserved, suggesting that there is an upper bound of the advancement of aging beyond which one cannot survive on average. The result further leads us to a statement: the product of the mean life span, $T$, and the mean rate of aging, $\bar{v}_A$, is constant, which is reminiscent of the empirical relationship (Rubner, 1908; Pearl, 1928; Cutler, 1983, 1991), $\tilde{T} \times \bar{v}_M \approx constant$, between the mean life span and the mean basal metabolic rate, $\bar{v}_M$, once discarded because of little evidence to support it and apparent contradiction between interspecies. In what follows, we approach these problems from a theoretical point of view.

## §2. Theoretical

Consider a society comprised of biologically identical $n(t)$ members with equal chronological age, $t$. Consider a deterministic stochastic process where members can only disappear, which is an example of the familiar decay model of population (Karlin and Taylor, 1981). To treat aging phenomena mathematically, let us introduce the assumption:

Assumption I: *The society is ideal, where members disappear (die) only due to aging, with no accidental deaths, no starvation, no infectious diseases, and so forth.*

Under this assumption, a relationship can be derived between the extent of the advancement of aging, $A$, and the continuation probability, $P$, of survival (Suematsu, 1995): Let $\psi$ represent a function. By Assumption I, the probability, $-\delta n/n$, of disappearance of members at age $t$ must directly be linked with the advancement of aging, $A$, per unit member. Thus it follows that

$$\delta A(t) = \psi(-\delta n/n). \tag{1}$$

With $\psi(0) = 0$ in mind, expand eq. (1) into the Taylor series

$$\delta A(t) = k(-\delta n/n) + \mathcal{O}(-\delta n/n), \tag{2}$$

and integrate the resultant series to yield

$$A(t) - A(0) = -k \log P(t), \tag{3}$$

with $P(t) = n(t)/n(0)$ being the continuation probability of survival. If we set the boundary condition, $A(0) = 0$ for $t = 0$, eq. (3) then reduces to

$$A = -k \log P, \tag{4}$$

where $k$ is a constant. Eq. (4) links the quantity, $A$, with the survival probability, $P$.

There is a useful way to look at the biological aspect of eq. (4). A partial advancement of aging from age $t_1$ to $t_2$ is written as

$$\Delta A_{21} = -k \log P(t_2)/P(t_1),$$

and the corresponding quantity from $t_2$ to $t_3$ is

$$\Delta A_{32} = -k \log P(t_3)/P(t_2).$$

The overall advancement of aging from age $t_1$ to $t_3$ then becomes

$$\Delta A_{31} = -k \log P(t_3)/P(t_1) \equiv \Delta A_{21} + \Delta A_{32}. \tag{5}$$

It turns out that the total advancement of aging is equal to the sum of the partial quantities. The quantity, $A$, is additive in nature, in accord with our experiences. We show in the following that some new relationships can be derived from this equality.

Let $t$ be an age and differentiate eq. (4) with respect to age $t$ to yield:

$$\dot{A} = k(-\dot{P}/P), \tag{6}$$

where the symbol *dot* denotes the differentiation with respect to age $t$. The term $(\cdots)$ of the right-hand-side is the familiar mortality function, $\mu$, of Gompertz. It turns out that the Gompertz function represents the measure of the rate of the advancement of aging. It has been well established that the mortality function is an increasing function of age (Pollard, 1973) and attains the maximum point, $\pi_0$, beyond the critical age, $t_c$ (Carey et al, 1992; Curtsinger et al, 1992; Rose, 1994; Manton et al, 1995; Suematsu, 1995, 1999; Vaupel et al, 1998; Wang et al, 1998). Thus we may have the statement:

The velocity of aging increases with advancing age

$$d\dot{A}/dt \geq 0, \tag{7}$$

and approaches the maximum speed at high age, or

$$\ddot{A} \to 0 \quad \text{for } t \geq t_c. \tag{8}$$

These can be explained reasonably in mathematical terms through the Taylor expansion of $\mu$ with respect to $P$; i.e.,

$$\mu(P) = -\dot{P}/P = \pi_0 + \pi_1 P + \pi_2 P^2 + \cdots, \tag{9}$$

where $\{\pi_k\}$ are coefficients. Observations have shown that $\mu$ is continuous and finite for all $P$ of $0 \leq P \leq 1$, suggesting that eq. (9) is generally convergent. Thus, for sufficiently high age, $P \to 0$, so that $\mu \to \pi_0$. The $t_c$ mentioned above is therefore an age above which $\mu = \pi_0$ is approximately satisfied (and does not necessarily mean a clear-cut critical age).

For the higher age where $\mu \cong \pi_0$ is satisfied, $\pi_0$ can be equated with the reciprocal of the mean life-span as discussed previously[†] (Suematsu, 1995, 1999).

— We can assess the physical soundness of eq. (4). For this purpose, in Fig. 1, $A$ is plotted as a function of age, $t$, where $P$ is assumed to follow the general sigmoid curve of advanced societies, but the scale is arbitrary since $k$ is indeterminable at present. As is seen in Fig. 1, $A$ increases, steeply at younger age, but linearly at higher age where $A \propto t$. It is widely accepted that the aging is a direct consequence of biological deterioration. If this is true, the biological function must decline progressively with advancing age, corresponding to the behavior of $A$. Recent observations show this trend: Wilson and coworkers could show that the cognitive ability declines more rapidly in older persons than in younger persons (Wilson et al, 1999), consistent with the behavior of eq. (4) (see Fig. 1). —

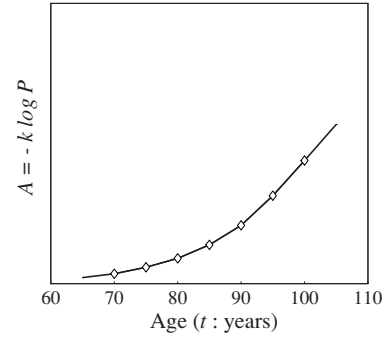

Fig. 1: The extent, $A$, of the advancement of aging as a function of age $t$.

The physical meaning of the constant $k$ of eq. (4) can be specified. Note that the death rate, $-\dot{P}$, satisfies the normalization condition in the interval $t = [t_0, \infty]$:

$$\int_{t_0}^{\infty} (-\dot{P}) dt = 1, \tag{10}$$

where $t_0 \ (\geq 0)$ is a point on a time axis that satisfies $P(t_0) = 1$; the interval $t = [0, t_0]$ is therefore defined as a period in one's life during which the extent of the advancement of aging remains 0, namely,

$$A(t) = 0 \text{ for } 0 \leq t \leq t_0.$$

Thus it follows that $t_0$ is not necessarily equal to 0 years.

Let $T(t)$ be the mean life-span from age $t$. With eq. (10), one immediately gets the expression for the mean life-span from age $t_0$:

$$T(t_0) = \int_{t_0}^{\infty} (t - t_0)(-\dot{P}) dt = \int_{t_0}^{\infty} P \, dt. \tag{11}$$

Since $-\dot{P}$ is normalized in the interval $-\dot{P}$ also, one can write

$$T(0) = \int_0^\infty t(-\dot{P})dt = \int_0^\infty P\,dt. \qquad (12)$$

The right-hand-side integral is separable, and

$$T(0) = \int_0^\infty P\,dt = \int_0^{t_0} P\,dt + \int_{t_0}^\infty P\,dt. \qquad (13)$$

According to the definition of $t_0$, $P=1$ for $0 \le t \le t_0$, and one has the equality:

$$T(0) = t_0 + T(t_0). \qquad (14)$$

Now let us derive some related functions of $A$. With $P = e^{-\beta A}$ ($\beta = 1/k$), the mth moment of $A$ is given by

$$\langle A^m \rangle = \int_0^\infty A^m(-\dot{P})dt, \qquad (15)$$

which, since $A^\ell P \to 0$ ($\ell = 0,1,\ldots,m$) for both $t \to 0$ and $\infty$, yields

$$\langle A^m \rangle = \frac{m!}{\beta^m}. \qquad (16)$$

The case of $m=1$ is of special interest.

$$\langle A \rangle = \frac{1}{\beta}\,(=k). \qquad (16')$$

Eq. (16') states that the mean extent of the advancement of aging throughout one's life is constant; conversely, this may be interpreted as: One can not continue to survive, on average, beyond the critical quantity, $A_c = \langle A \rangle\,(=k)$, which, following the relationship of eq. (4), corresponds to

$$P = e^{-\beta\langle A \rangle} = e^{-1} \cong 0.37. \qquad (17)$$

This is the ideal limit of the mean survival probability based on $A_c$.

Let us apply this result to the $P$ expansion of the mortality function (Suematsu, 1995, 1999). One has for sufficiently high age

$$p(t) = \exp(-\pi_0 t), \quad \text{for } t \ge t_c \qquad (18)$$

where $p(t)$ is the normalized survival probability, $p(t) = P(t)/P(t_c)$, and $t_c$ the critical age above which eq. (18) applies. We can now transpose the same aging quantity to this regime, and thus

$$A = -k\log p. \qquad (4')$$

The mean extent of the advancement of aging of this regime is again $\langle A \rangle = k$. At $A = \langle A \rangle$, one has

$$p(T) = e^{-1}. \qquad (17')$$

Comparing with eqs. (17') and (18), one finds

$$T = 1/\pi_0 \text{ for } t \ge t_c,$$

which is exactly the previous result: The mean life-span is constant for higher ages of $t \ge t_c$. This is an example that the mean life span based on the aging concept coincides with $T$ defined by eq. (11).

From eq. (16') we get an implication that since one cannot survive, on average, beyond $A_c$, all one can do for longevity is only to delay the advancement of aging.

The statement can be made clearer by transforming eq. (16') as

$$T\frac{\langle A \rangle}{T} = k, \qquad (19)$$

The second term $\langle A \rangle/T$ of the left-hand-side represents the velocity, $\bar{v}_A$, of aging averaged out throughout one's life span. Thus,

$$T \times \bar{v}_A = k, \qquad (20)$$

$k$ being the mean extent of the advancement of aging as mentioned above. Now one has the theorems:
(*1*) total quantity of aging throughout one's life is constant, on average;
(*2*) the product of $T$ and $\bar{v}_A$ is constant.
$T$ is in inverse proportion to $\bar{v}_A$. The lesser the velocity of the advancement of aging, the longer the life span.

For the higher age of $t \ge t_c$, $T = 1/\pi_0$, so that $\bar{v}_A$ is a constant, $k\pi_0$, independent of age. The recent works by Hwang, Krapivsky and Redner (Hwang et al, 1999, 2002) deal with the solution of the conservation equation for this regime with the aging concept consistent with the present work.

The essence of the above result is represented in Fig. 2 where the age-change of $v_A$ is shown for different two cohorts (*a*) and (*b*); the x-axis being $t$ and the y-axis $v_A$. The representation may be viewed as an example of annual change of the mortality function observed commonly in advanced societies; the curve (*a*) representing an old generation and (*b*) a younger generation. The areas enclosed by curve (*a*)-$t_0$-$t_1$ and curve (*b*)-$t_0$-$t_2$ stand for total quantities of aging, $\langle A \rangle$, of the corresponding generations. According to the theorem (*1*), the two areas are equal, so that $t_1$ corresponds to the mean life-span of the cohort (*a*) and $t_2$ that of (*b*), respectively. Thus the areas of the two shaded loops must be equal. It is seen that the cohort (*b*) has lower $v_A$ than (*a*) throughout the whole life-span and therefore has a longer mean life-span, $t_2 > t_1$. By the shift from the generation (*a*) to (*b*), the mean life span increased by $\Delta t = t_2 - t_1$.

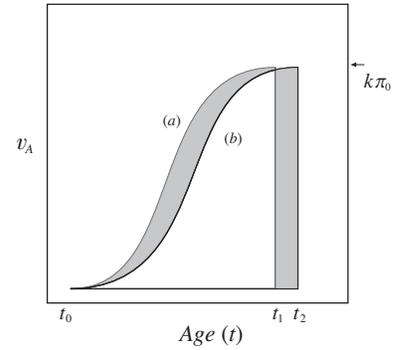

Fig. 2: Representation of the shift of $v_A$ vs $t$ curves from an old to a younger generation. $v_A$ was drawn using recent observed data of the mortality function, $\mu$.

From a biological point of view, $T$ in eq. (20) should be equated with $T(t_0)$, since $A = 0$ for $0 \le t \le t_0$ and aging cannot be defined for the interval, $0 \le t \le t_0$ (Suematsu, 1995); $t_0$ is an age where aging starts to advance. Unfortunately, until today no theoretical and experimental verifications for $t_0$ have not been put forward. Here we replace, for practical purposes, $T$ with $T(0)$. Rigorously, this replacement might be controversial, but useful for the purpose of testing the validity of eq. (20).

Eq. (20) states that only one possible way to elongate one's life span is to lower the velocity, $\bar{v}_A$, of aging. Note that eq. (20) does not specify a path leading to the critical quantity, $A_c$; i.e., it does not preclude the possibility that one can change $\bar{v}_A$. If this is true, eq. (20) may provide us with a possible strategy to elongate one's life span. Then let us assess the applicability of eq. (20).

### §3. COMPARISON WITH EXPERIMENTS
### 3-1). AGING IN CELLS

Aging should be a result of the decline of biological functions. According to Assumption I, the decline of the functions can be related to the disappearance of a member —Replacing the 'biological' with 'mechanical' and the 'disappearance of a member' with 'failure of products', this definition of aging may be applied to more general cases, including the aging of industrial materials (Barlow and Proschan, 1981; Koltover, 1982, 1997) —.

Let us focus our attention on the problem of cellular aging. It appears obvious that the decline of the cell replicability is closely related to the decline of the biological functions. Thus, it will be reasonable to generalize the basic concept of aging to include the aging of cell divisions.

It has been well established that cancer cells, typically the Hela cell, exhibit no appreciable aging in their replicability in contrast to ordinary body cells (Avernathy, 1998). To be specific, it has been observed that the period of the cell division of the Hela cell does not change with advancing age; they show no appreciable aging in the replication rate. This can be regarded as a typical example of $\bar{v}_A \approx 0$. Substituting this into eq. (20), one has $T \approx \infty$. The result is in good accord with our observations, the unlimited replicability of tumor cells.

### 3-2) HIBERNATING HAMSTERS

It was found that hibernating hamsters live longer than those that don't hibernate. This phenomenon was first established by Lyman and coworkers, who coined the term, 'suspended animation' (Lyman et al, 1981). Following Lyman and coworkers, let us accept the notion of the suspended animation. Then the hibernation should directly lead to the lowering of the rate of aging. Let $\chi$ be the fraction of the total hibernating period throughout a whole life span. Following the aforementioned additive law of aging, the mean critical extent of aging must be a sum of the mean extent of aging during the non-hibernating periods and that during the hibernating periods:

$$\langle A \rangle = \langle A_N \rangle + \langle A_H \rangle = T\{\bar{v}_{A,N}(1-\chi) + \bar{v}_{A,H}\chi\}, \quad (21)$$

with the subscripts $H$ and $N$ denoting the hibernation and the non-hibernation, respectively. The general expression of the mean rate of aging is therefore given by

$$\bar{v}_A = \bar{v}_{A,N}(1 - \chi + g\chi), \quad (22)$$

where $g = \bar{v}_{A,H}/\bar{v}_{A,N}$. No rigorous calculation of $g$ is available at present, but an approximate form of eq. (22) can be derived: By the assumption of the suspended animation, it follows that $0 \leq g \leq 1$. We are interested in the case, $g \ll 1$, then

$$\bar{v}_A \cong \bar{v}_{A,N}(1 - \chi). \quad (23)$$

Substituting this into eq. (20), we have

$$T \cong \left(k/\bar{v}_{A,N}\right)\frac{1}{1-\chi} \quad \text{for } g \ll 1, \quad (24)$$

the prefactor $(k/\bar{v}_{A,N})$ being experimentally determinable as a limiting case of $\chi = 0$. We can now compare the present theory with the Lyman and co-workers data (Lyman et al, 1981). Prior to comparison, it is convenient to average out the experimental points every narrow interval of $\chi$ ($\Delta\chi \leq 0.005$ for Fig. 3-a; $\Delta\chi = 0.05$ for Fig. 3-b). Numerical data ($\diamond$ and $\times$) thus obtained are plotted in Fig. 3 together with the theoretical equation (24) (solid line) which is given the prefactor, $(k/\bar{v}_{A,N}) = 750$. As one can see, agreement between the theory and the experiment is excellent, in support of the mathematical soundness of eq. (20); the result giving a confirmation that the hibernation lowers the rate of aging.

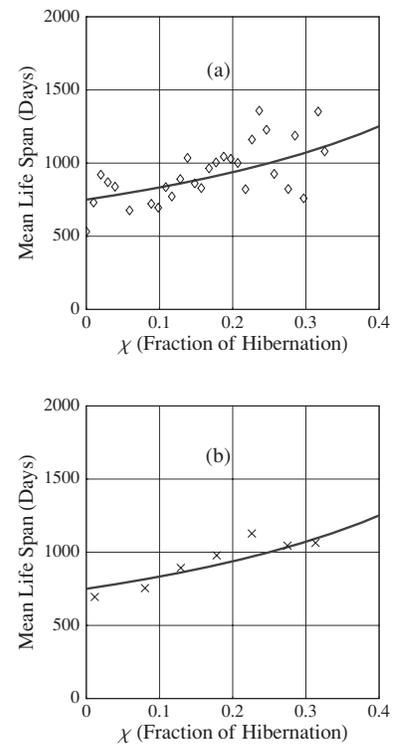

Fig. 3: Mean life span (days) as a function of the fraction ($\chi$) of the total hibernation. Solid line: eq. (24). Observed points: $\diamond$ and $\times$ by Lyman, O'Brien, Greene and Papafrangos. The original data were averaged in the interval, (a) $\Delta\chi \leq 0.005$ ($\diamond$), (b) $\Delta\chi \leq 0.05$ ($\times$).

### 3-3). MAMMALIAN AGING

In 1780, Lavoisier and Laplace carried out the first quantitative investigation of mammalian activity. They measured heat evolved by a guinea pig placed in the calorimeter and compared it with the amount of oxygen consumed by the animal. After close investigation, they reached the conclusion that respiration is a slow combustion (Moor, 1972; Kleiber, 1975). As plenty of data have been accumulated since Lavoisier's days, it became recognized that basal metabolic rates per unit mass (BMR), $\bar{v}_M \left[KJoules\ Kg^{-1}\ Day^{-1}\right]$, are intimately related with life-spans of mammals. According to the findings, BMR's are roughly in inverse proportion to life-spans between interspecies for most mammals; that is to say, total amounts of metabolism per unit mass, $\langle M \rangle\ \left[KJoules\ Kg^{-1}\right]$, throughout their lives are roughly constant irrespective of species, which leading to the empirical relationship:

$$\tilde{T} \times \bar{v}_M \approx constant. \quad (25)$$

where $\tilde{T}$ denotes an age when 90% mortality occurs (Cutler, 1983, 1991). Although $\tilde{T}$ has no direct connection with the above-defined statistical quantity, $T$, it may be regarded roughly to represent the mean life span. The finding has lead biologists to the notion of the rate-of-living hypothesis (Rubner, 1908; Pearl, 1928; Kleiber, 1975). The historical background can be found in the review articles by Heusner (Heusner, 1985), and Holliday, Potter, Jarrah, and Berg (Holliday et al, 1967). Observed data from numerous authors were summarized by Cutler (Cutler, 1983, 1991: see Fig. 4) who compared BMR's and average ages, $\tilde{T}$, of various mammals over wide interspecies.

In Fig. 4, an example of the theoretical line given a value of *constant* = $2\times10^6$ (solid line) is shown to compare with the observed points ( : primates; : African elephant, horses, and red deer; : rodents). The horizontal axis indicates BMR $\left[KJoules\ Kg^{-1} Year^{-1}\right]$ and the vertical axis the life span, $\tilde{T}$. Agreement of the empirical relationship (25) and the observations is satisfactory, which may lead to the correspondence:

$$\bar{v}_A \leftrightarrow \bar{v}_M$$
$$\langle A \rangle \leftrightarrow \langle M \rangle,$$

namely,

$$T \times \bar{v}_M = k_M, \qquad (26)$$

where $k_M$ is a constant.

Eq. (26) implies that the mean basal metabolic rate is in inverse proportion to the mean life span as has been suggested earlier. It appears from eqs. (20) and (26) that the mean aging rate correlates with the mean basal metabolic rate, and the total extent of the advancement of aging the total metabolism per unit mass. The advancement of aging appears deeply connected with the basal metabolism.

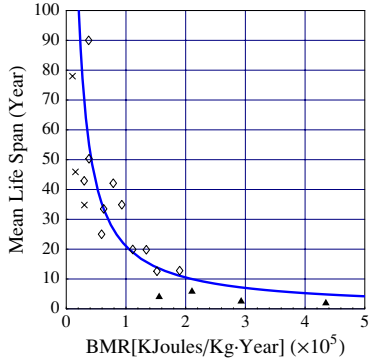

Fig. 4: Relation of mean life span $T$ [Year] and basal metabolic rate [K Joules Kg$^{-1}$Year$^{-1}$].

BMR's were evaluated as quantities per Year rather than Day in order to conform to the unit of corresponding mean life spans [Year]. Thus the total metabolism is evaluated (from an example of a human) to be $T \times \bar{v}_M \approx 55 \cdot 100 \cdot 365 \approx 2 \times 10^6$ [K Joules Kg$^{-1}$]. Solid line: eq. (26), $k_M = 2 \times 10^6$ [K Joules Kg$^{-1}$]. Observed data (quoted from Cutler, 1983): : primates (man, orangutan, gorilla, capuchin, rhesus, lemur, gibbon, owl monkey, marmoset, tree shrew, tarsier); : African elephant, horses, and red deer; : rodents (rat, gerbil, mouse, pygmy shrew).

### 3-3-1) REASON OF DIFFERENCE IN LIFE SPAN BETWEEN WOMEN AND MEN

Let us accept, for a while, the commensuration of eqs. (20) and (26), and test the applicability to observations. According to the relationship (26), the relative mean life span is expected to be inversely proportional to the relative mean metabolic rate. A rough correlation has been established for interspecies as mentioned above. According to Assumption I, however, the relation (26) must be verified rigorously using biologically identical samples. The most reliable data will be those for humans: Human BMR's were measured early in the last century by Harris and Benedict (Harris and Benedict, 1919; Benedict, 1928, 1938). Mean metabolic rates[†] are readily accessible from these data, and calculated to be

$$\bar{v}_M \cong \begin{cases} 98.5\ \left[KJoules\ Kg^{-1}\ Day^{-1}\right]\ for\ men \\ 91.2\ \left[KJoules\ Kg^{-1}\ Day^{-1}\right]\ for\ women \end{cases}, \qquad (27)$$

respectively. According to the relationships (20), (26), we expect that the women in question can live longer 1.08 times than the men can do. Namely, we expect

$$\frac{T_{Women}}{T_{Men}} = \frac{98.5}{91.2} \cong 1.08. \qquad (28)$$

Unfortunately, no information is available for the life spans of those women and men (Harris and Benedict, 1919; Benedict, 1928, 1938). So, we examine eq. (28) with recent observations based on population statistics. The annual data of the mean life span of Japanese women and men from 1920 to 1997 (Yasuhara, 1994) are plotted in Fig. 5. The solid line is the linear equation (28), $T_{Women} \cong 1.08 \cdot T_{Men}$, expected from the relative metabolic rate. The symbols, , denote the observed data of the relative mean life span of women to men in Japan.

Agreement between the predicted line (28) and the observed data is excellent. To confirm this result, eq. (28) is compared with statistics of other nations ( : Sweden, : Netherlands, : Switzerland, : Norway) from 1967 to 1997 (Yasuhara, 1994). The results are shown in Fig. 6. Agreement between the observed data and the expected line is again excellent.

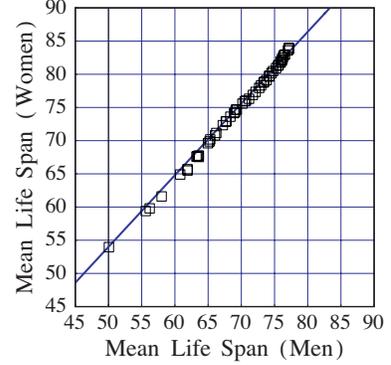

Fig. 5. Relative mean life span of women to men.
Solid line: theoretical, $T_{Women} \cong 1.08 \cdot T_{Men}$, eq. (28). : observed data from 1920 to 1997 in Japan.

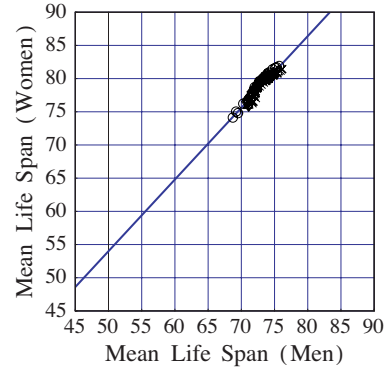

Fig. 6. Relative mean life span of women to men.
Solid line: theoretical, $T_{Women} \cong 1.08 \cdot T_{Men}$, eq. (28). Observed data: : Sweden; : Netherlands; : Switzerland; : Norway, from 1967 to 1997.

[†] Note that

$$\bar{v}_A = \frac{\int_0^\infty A(-\dot{P})dt}{\int_0^\infty t(-\dot{P})dt} = \frac{\int_0^\infty \dot{A}P\,dt}{\int_0^\infty P\,dt} = \int_0^\infty v_A q(t)dt,$$

and thereby $q(t)$ expresses the probability of a member lying at age $t$. Likewise, $\bar{v}_M$ is defined by $\bar{v}_M = \int_0^\infty v_M q(t)dt$. For a practical purpose, $\bar{v}_M$ is calculated by multiplying observed $v_M$ by the weight $q(t)$ and summig over all age.

At first sight, the unexpectedly good agreement between eq. (28) and the observations in Figs. 5 and 6 is disturbing. The mean life span is at present rapidly increasing with year. It, thus, follows that the mean rate of aging, $\bar{v}_A$, must be decreasing with year.

An immediate intuition will be that the relative mean life span should asymptotically approach to a given value, say, 1.08 with year, however the results of Figs. 5 and 6 show that this is not the case. One explanation for the remarkable agreement is that the mean rate of aging, $\bar{v}_A$, is in fact now decreasing with year, while the relative value, $\bar{v}_{A,Men}/\bar{v}_{A,Women}$, remains invariant. This is by no means an unreasonable explanation. Recent observed data show that the decrease of $\bar{v}_A$ is, in fact, what is occurring (Yasuhara, 1994): the quantity, $\mu$, equivalent to $v_A$ is now rapidly decreasing with year over all generations from 0 to 100 years of age, consistent with the above results. In addition, some fraction of the decrease of $\bar{v}_A$ may come from the decrease of the death from essential diseases (not accidental) and the death from those diseases approximately follows the Gompertz function, namely $v_A$ itself (Kohn, 1978). The change of $v_A$ is indistinguishable from that due to the death from essential diseases.

### 3-3-2) THE CASE OF RATS

BMR's of the laboratory rats, were measured by Benedict (Benedict, 1928, 1938). According to his paper, the ages of the rats investigated ranged from 180 to 660 days, and the measurement was made near 28 $^\circ C$. The numerical values can be read from his figure (some uncertainty may therefore arise), which gives

$$\frac{\bar{v}_{M,Male}}{\bar{v}_{M,Female}} \approx 0.9. \qquad (29)$$

Thus we expect the relative mean life span of female rats to males as

$$\frac{T_{Female}}{T_{Male}} \approx 0.9 \text{ for rats}. \qquad (30)$$

In contrast to the case of humans, male rats should live longer than females.

Let us examine our expectation by observations. The relationship between the mean life spans of female and male rats have been investigated by Hursch et al using Wister rats of an identical strain (Hursch et al, 1955). They showed

$$\frac{T_{Female}}{T_{Male}} \approx 0.85. \qquad (31)$$

Taking into account that their observed data were made with small sample size ($N = 23$) so that the numerical values are not very reliable, agreement of eq. (30) with eq. (31) is satisfactory. The general trend, the inverse proportionality between mean life spans (mean rates of aging) and mean metabolic rates,

$$\frac{T_{Female}}{T_{Male}} \equiv \frac{\bar{v}_{A,Male}}{\bar{v}_{A,Female}} \sim \frac{\bar{v}_{M,Male}}{\bar{v}_{M,Female}}, \qquad (32)$$

is clearly manifested (the symbol $\sim$ denotes an assumed equality to be proven); male rats live longer than females, as expected from eqs. (31) and (32).

### §4. DISCUSSION

For the examples taken up in this article, the theory agrees well with the experimental observations; no inconsistency can be found, in support of the physical soundness of eqs. (4) and (20): It was found that eq. (20) is in harmony with our observations for the unlimited replicability of tumor cells and the prolonged life spans of hibernating hamsters (§ 3-1 and 3-2); eq. (20) is consistent with the recent observations in population statistics, the increasing mean life span and the decreasing mean rate of aging (Gompertz function).

Whereas eq. (20) is on the firm foundation, the empirical relationship (26) remains conjecture. We discuss this point briefly:

It appears that the empirical equation (26) explains well the difference in life span between male and female of mammals, but the theory (20) implies that the relation (26) must be treated with caution: It can not explain the unlimited replicability of tumor cells. Recall that $T \approx \infty$ for tumor cells, so that if the relation (26) is true, they must have zero metabolic rate, $\bar{v}_M = 0$, which contradicts our experiences. Moreover, it has been found that the mean basal metabolic rate gradually decreases with age (Benedict, 1928, 1938), while the rate of aging equivalent to the mortality function, $\mu(t)$, increases steeply with advancing age (Pollard, 1973; Wilson, 1993; Suematsu, 1995, 1999). It turns out that the rate of aging is not identical with the metabolic rate.

Notwithstanding, the results of § 3-3 appear to suggest strongly that the respective mean values, $\bar{v}_A$ and $\bar{v}_M$, are deeply connected with each other. It is possible that the relation (26) is true for the aging of a body such as the mammals. According to eq. (20), the ratio of the mean rate of aging can be equated with the inverse ratio of the mean life span, namely, $\frac{T_{Female}}{T_{Male}} = \frac{\bar{v}_{A,Male}}{\bar{v}_{A,Female}}$, this ratio being conserved as an invariant quantity, while the results of Figs. (5) and (6) suggest that these may be intimately connected with the corresponding ratio of the mean basal metabolic rate, $\bar{v}_{M,Male}/\bar{v}_{M,Female}$, as an invariant quantity as well. This leads us to the assumed equality (32). Whether eq. (32) is universally valid for the aging of a body remains conjecture. To prove this, it is necessary to confirm that the mean metabolic rate, $\bar{v}_M$, is, in fact, decreasing with year in parallel with $\bar{v}_A$ so that the relation (26) can account for the recent increase of the mean life span. It is by no means unlikely that $\bar{v}_A$ is now decreasing with year, say in Japan, if we recall the rapid increase of the body size of Japanese people. Indeed, one example in USA seems to support this conjecture: In Table 1 and Fig. 7, the data in 1919 by Harris and Benedict is compared with their own data in 1928 (Harris and Benedict, 1919; Benedict, 1928, 1938).

Table 1. Annual change of $\bar{v}_M$

|  | $\bar{v}_{M,Men}$ [$KJoules\ Kg^{-1}\ Day^{-1}$] | $\bar{v}_{M,Women}$ [$KJoules\ Kg^{-1}\ Day^{-1}$] |
|---|---|---|
| Data in 1919 | 108 | 102 |
| Data in 1928 | 98.5 | 91.2 |

It is found that during the decade from 1919 to 1928, the mean metabolic rate decreased to $\bar{v}_A \cong 90/100$ for both men and women (it is worth noting that the decrease of $\bar{v}_M$ occurs over all generations; see Fig. 7), while the mean life span in USA in the same period is reported to have increased by $\cong 1.06$ times for men and $\cong 1.08$ times for women, respectively (Yasuhara, 1994). These seem to well satisfy the relation (26), in support of the annual change of $\bar{v}_M$ and the soundness of the reciprocal relation (32).

In order to lay the relation (26) on the firm foundation, however, we must scrutinize the annual change of $\bar{v}_M$ with more extensive data of population statistics. To date, no such experiment has been carried out. Whether the remarkable agreement seen in Figs. (5) and (6) is a natural consequence of the reciprocal relation (32) depends on whether the annual change of $\bar{v}_M$ can be confirmed. If this should be refuted, then the source of the mysterious value, 1.08, must be sought to the others.

On the other hand, if this is in fact confirmed, then eq. (32) is valid, and so is the relation (26), and the aforementioned factor $g$ can be equated with the relative mean metabolic rate of hibernation to non-hibernation. Then we will be able to accomplish a complete plot of the suspended animation according to the general expression:

$$T = \left(k/\bar{v}_{A,N}\right)\frac{1}{1-\chi+g\chi}. \qquad (24')$$

To this end, it is essential that the relation (32) is examined by extensive experimental observations.

## §5. Conclusion

1. The equality between the extent of the advancement of aging and the survival probability,

$$A = -k\log P, \qquad (4)$$

leads to a consequence that the total advancement of aging throughout one's life is conserved:

$$\langle A \rangle = k. \qquad (16')$$

From eq. (16'), we have a theorem: the product of the mean life span, $T$, and the mean rate of aging, $\bar{v}_A$, is constant

$$T \times \bar{v}_A = k. \qquad (20)$$

The lesser the velocity, $\bar{v}_A$, of the advancement of aging, the longer the life span. The result is in harmony with our experiences: the unlimited replicability of Tumor cells, $T \approx \infty$ for $\bar{v}_A \approx 0$; the increasing mean life span and the decreasing mean rate of aging. If we accept the notion of the suspended animation, eq. (16) can be generalized to account for the mean life span of hibernating animals

$$T \cong \left(k/\bar{v}_{A,N}\right)\frac{1}{1-\chi}, \qquad (24)$$

where $\chi$ is the fraction of hibernation throughout a whole life length. Eq. (24) is in good accord with the Lyman and coworkers observations in Turkish hamsters (Fig. 3).

2. Eq. (20) yields, in a natural fashion, $\frac{T_{Female}}{T_{Male}} = \frac{\bar{v}_{A,Male}}{\bar{v}_{A,Female}}$. The relative mean life span of female to male is inversely proportional to the corresponding relative value of the mean rate of aging. The results of § 3-3 suggest that these quantities may be connected with the relative mean metabolic rate:

$$\frac{T_{Female}}{T_{Male}} = \frac{\bar{v}_{A,Male}}{\bar{v}_{A,Female}} \sim \frac{\bar{v}_{M,Male}}{\bar{v}_{M,Female}}. \qquad (32)$$

To prove this, it is necessary that the annual change of the basal metabolic rate, $\bar{v}_M$, can be experimentally confirmed; one example in USA seems to support this.

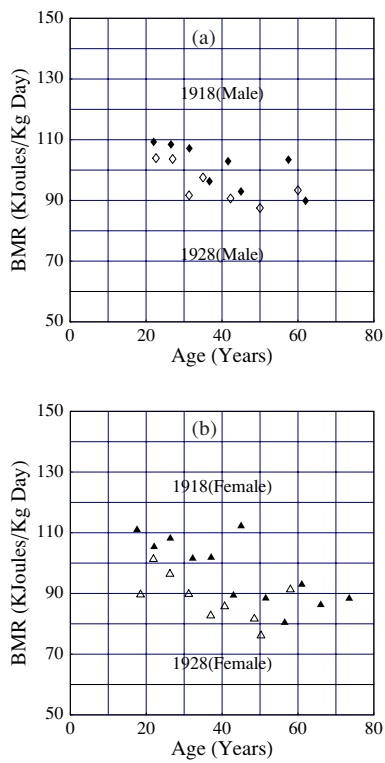

Fig. 7. Annual change of $\bar{v}_M$.
The original data by Harris and Benedict were averaged every interval, $\Delta t \leq 5$, and plotted as against age $t$: (a) Men (  ) in 1919; (  ) in 1928. (b) Women (  ) in 1919; (  ) in 1928.

## References

Avernethy, J. 1998. Gompertzian Mortality Originates in the Winding-down of the Mitotic Clock. J. Theor. Biol. 192, 419-435.

Barlow, R. E. and Proschan, F. 1981.
Statistical Theory of Reliability and Life Testing. To Begin With, Silver Spring, MD.

Benedict, F. G. 1928. Basal Metabolism Data.
Am. J. Physiol., 85, 607.

Benedict, F. G. 1938. Vital Energetics: A Study in Comparative Basal metabolism. Carnegie Institution of Washington, Publication No. 503, 93.

Carey, J. R., Liedo, P., Orozco, D. and Vaupel, J. W. 1992. Slowing of Mortality Rates at Older Ages in Large Medfly Cohorts. Science, 258, 457-460.

Curtsinger, J. W., Fukui, H. H., Townsend, D. R. and Vaupel, J. W. 1992.
Demography Genotypes: Failure of the Limited Life-Span Paradigm in Drosophila melanogaster. Science, 258, 461-463.

Cutler, R. G. 1983.
Superoxide Dismutase, Longevity and Specific Metabolic Rate. Gerontology, 29, 113.

Cutler, R. G. 1991.
Human Longevity and Aging: Possible Role of Reactive Oxygen Species. Annals New York Academy of Sciences, 621, 1.

Dodds, P. S. and Rothman, D. H. 2001.
Geometry of river networks. Phys. Rev. E, 63, 016115; 2001; Re-examination of the "3/4-law" of Metabolism. J. Theor. Biol., 209, 9.

Harris, J. A. and Benedict, F. G. 1919.
A Biometric Study of Basal Metabolism in Man. Carnegie Institution of Washington, Publication No. 279, 1.

Heusner, A. A. 1985.
Body Size and Metabolism. Ann. Rev. Nutr., 5, 267.

Holliday, M. A., Potter, D., Jarrah, A., and Berg, S. 1967. The Relation of Metabolic Rate to Body Weight and Organ Size. Pediat. Res., 1, 185.

Hursch, J. B., Noonam, T. R., and Casarett, G. 1955.
Reduction of Life Span of Rats by Roentgen Irradiation. Amer. J. Roentgenol., 74, 130.

Hwang, W., Krapivsky, P. L. and Redner, S. 1999. Does Good Mutation Help You Live Longer? Phys. Rev. Letters, 83, 1251.

Hwang, W., Krapivsky, P. L. and Redner, S. 2002.
Fitness versus longevity in age-structured population dynamics. J. Math. Biol., 44, 375.

Karlin, S. and Taylor, H. M. 1981.
A second course in stochastic processes, New York ; London : Academic Press.

Kleiber, M. 1975.
The Fire of Life: An Introduction to Animal Energetics. Robert E. Krieger Pub. Comp. Malabar, Florida.

Kohn, R. R. 1978.
Foundations of Developmental Biology Series: Principles of Mammalian Aging. Prentice-Hall, Inc., Englewood Cliffs, New Jersey, USA.

Koltover, V. K. 1982.
Reliability of enzyme systems and molecular mechanisms of ageing. Biophysics, 27, 635.

Koltover, V. K. 1997.
Reliability Concept as a Trend in Biophysics of Aging. J. Theor. Biol., 184, 147.

Lyman, C. P., O'Brien, R. C., Greene, G. C. and Papafrangos, E. D. 1981.
Hibernation and Longevity in the Turkish Hamster. Science, 212, 668.

Manton, K. G. and Vaupel, J. W. 1995.
Survival after the Age of 80 in the United States, France, England, and Japan. The New England Journal of Medicine, 333, 1232, 1235.

Medvedev, Z. A. 1990.
An Attempt at a Rational Classification of Theories of Aging. Biol. Rev., 65, 375-398.

Moor, W. J. 1972.
Physical Chemistry. Prentice-Hall, Inc., Englewood Cliffs, New Jersey, USA.

Pearl, R. 1928.
The Rate of Living. New York, Alfred A. Knopf.

Pollard, J. H. 1973.
Mathematical models for the growth of human populations. Cambridge, Cambridge University Press.

Rose, M. R. 1994.
Evolutionary Biology of Aging. New York, Oxford Univ. Press.

Rubner, M. 1908.
Das Problem der Lebensdauer und seine Beziehungen zum Wachstum und Ernahrung. Munich, Oldenbourg.

Sacher, G. A. 1982. Evolutionary theory in gerontology. Perspectives in Biology and Medicine, 25, 339-353.

Sacher, G. A. 1967.
The complementality of entropy terms for the temperature-dependence of development and aging. Annals of the New York Academy of Sciences, 138, 680-712.

Sacher, G. A. 1980.
Theory in gerontology, part I. Annual Reviews of Gerontology and Geriatrics, 1, 3-25.

Strehler, B. L. 1967.
Cellular Aging. Annals of the New York Academy of Sciences, 138, 661.

Suematsu, K. 1995.
A Note on Population Dynamics. J. Theor. Biol., 175, 317.

Suematsu, K. 1999.
Age Invariant of Gompertz Function and Exponential Decay of Population. J. Theor. Biol., 201, 231.

Suematsu, K. (23, Dec) 2003.
http://arxiv.org/abs/q-bio.PE/0312037.

Vaupel, J. W., Carey, J. R., Christensen, K., Johnson, T. E., Yashin, A. I., Holm, N. V., Iachine, I. A., Kannisto, V., Khazaeli, A. A., Liedo, P., Longo, V. D., Zeng, Y., Manton, K. G., and Curtsinger, J. W. 1998.
Biodemographic Trajectories of Longevity. Science, 280, 855-860.

Wang, S., Matsushita, T., Kogishi, H., Xia, C., Chiba, T., Hosokawa M. and Higuchi, K. 1998.
Type B apoA-II and SAM: Biomedical Gerontology, 22, No1, 26.

Wilson, D. L. 1993.
A Comparison of Methods for Estimating Mortality Parameters: from Survival Data. Mechanism of Ageing and Development, 66, 269-281.

Wilson, R. S, Beckett, L. A., Bennett, D. A., Albert, M. S., and Evans, D. A. 1999.
Change in cognitive function in older persons from a community population: Archives of Neurology, 56, 1274-1279.

Yasuhara, T. 1994.
The Life Tables of Centenarians in Japan. Planning Division of Welfare for the Elderly, Minister's Secretariat, Japan.